\documentclass[12pt,preprint]{aastex6}
\usepackage{graphicx}
\usepackage{amsmath,amssymb}

\slugcomment{Accepted for publication in the Astrophysical Journal (11 November 2016)}

\begin{document}

\title{Sunspot and starspot lifetimes in a turbulent erosion model}

\author{Yuri E. Litvinenko} 
\affil{Department of Mathematics, University of Waikato, 
  P. B. 3105, Hamilton, New Zealand}
\and 
\author{M. S. Wheatland} 
\affil{Sydney Institute for Astronomy, School of Physics, 
  The University of Sydney, NSW 2006, Australia}

\begin{abstract}
Quantitative models of sunspot and starspot decay 
predict the timescale of magnetic diffusion and may yield 
important constraints in stellar dynamo models. 
Motivated by recent measurements of starspot lifetimes, 
we investigate the disintegration of a magnetic flux tube by nonlinear diffusion. 
Previous theoretical studies are extended by considering two physically motivated 
functional forms for the nonlinear diffusion coefficient $D$: 
an inverse power-law dependence $D \propto B^{-\nu}$ 
and a step-function dependence of $D$ on the magnetic field magnitude $B$.
Analytical self-similar solutions are presented for the power-law case, 
including solutions exhibiting ``superfast'' diffusion. 
For the step-function case, the heat-balance integral method yields  
approximate solutions, valid for moderately suppressed diffusion in the spot. 
The accuracy of the resulting solutions is confirmed numerically, using a method
which provides an accurate description of long-time evolution by imposing boundary 
conditions at infinite distance from the spot. The new models may allow insight 
into differences and similarities between sunspots and starspots.
\end{abstract} 


\section{Introduction}

A great deal of effort has gone into observing and analyzing disintegration 
of sunspots (and starspots). The sunspot decay is usually characterized 
by the rate of decrease of the sunspot area $A$, and numerous observations 
appear to be consistent with a parabolic decay law, with $A(t)$ 
a decreasing quadratic function of time $t$ 
\citep[e.g.][]{1988A&A...205..289M,1993A&A...274..521M,1997SoPh..176..249P}. 

Early theories invoked turbulent diffusion of the magnetic field 
within the spot to model the observed rate of decay, yet such models predicted 
a linear decay law, corresponding to a constant area decay rate $dA/dt$ 
\citep{1974MNRAS.169...35M,1975SoPh...42..107K}. 
In order to explain a parabolic decay, \citet{1997ApJ...485..398P}  
developed a model of sunspot disintegration by turbulent ``erosion'' 
of penumbral boundaries, which occurs when bits of magnetic field 
are sliced away from the edge of a sunspot and swept to the supergranular 
cell boundaries by supergranular flows \citep{1964ApJ...140.1120S}. 

A key feature of the erosion model is that the turbulent diffusivity, 
associated with the flows, is suppressed within the spot. 
The assumption is justified by the theoretical prediction that 
the diffusivity $D$ should rapidly decrease if 
the magnetic field $B$ exceeds an energy equipartition value 
\citep{1994AN....315..157K,2000AN....321...75R}, which is why 
the diffusivity in the turbulent erosion model may be assumed 
to be a decreasing function of the magnetic field strength. 
\citet{1997SoPh..176..249P} presented observational evidence 
in support of the parabolic decay law and its theoretical explanation 
by turbulent erosion. 

\citet{2015ApJ...800..130L} recently revisited the theory 
of sunspot decay by turbulent erosion, considered as a moving boundary problem. 
While some of the earlier results were confirmed for moderate 
sunspot magnetic field strengths, the new analytical and numerical solutions 
yielded a significantly improved theoretical description 
of sunspot disintegration. In particular, the dependence of the spot area 
was shown to be a nonlinear function of time, which in a certain parameter 
regime can be approximated by a parabola. More accurate expressions 
for the spot lifetime in terms of an initial magnetic field were derived 
analytically and verified numerically. 

Following \citet{1997ApJ...485..398P}, 
\citet{2015ApJ...800..130L} assumed in their study that the turbulent diffusivity 
$D=D(B)$ within a decaying sunspot is much less than that outside the spot. 
A more realistic model should incorporate a more realistic dependence 
of the turbulent diffusivity on the field strength within the spot. 
Our aim in this paper is further to develop the theory of turbulent erosion 
by exploring the effect of a non-vanishing diffusivity within a sunspot 
on the rate of its disintegration. 

Theoretical mechanisms and observable features of the diffusive transport of 
the photospheric magnetic field have been a subject of intense research activity 
\citep[e.g.][]{2006A&A...449..781C,2011ApJ...731L..39L,2012ApJ...759L..17L,2014ApJ...785...90R}. 
Our detailed analysis of a simple nonlinear model, 
reinforced by numerical solutions, can complement 
more detailed magnetohydrodynamic simulations 
\citep[e.g.][]{2008ApJ...684L.123H,2014ApJ...785...90R} 
and guide empirical models \citep{2014SoPh..289.1531G} 
in studies of sunspot and starspot evolution. 
The determination of lifetimes of spots is a topic 
of general---and current---astrophysical interest. 
Quantitative models for starspot evolution may help estimate 
the magnetic diffusion timescale and thus yield 
important constraints in stellar dynamo models 
\citep[e.g.][]{2014ApJ...795...79B,2015ApJ...806..212D,2015A&A...578A.101K}, 
which provides further motivation for our exploration of the turbulent erosion model.

\section{Formulation of the problem}

Following \citet{1997ApJ...485..398P} and \citet{2015ApJ...800..130L}, 
we model a decaying sunspot as a cylindrically symmetric flux tube. 
The evolution of the magnetic field ${\bf B} = B(r, t) \hat{\bf z}$ 
is governed by the following nonlinear diffusion equation: 
\begin{equation}
  \frac{\partial B}{\partial t} = 
  \frac{1}{r} \frac{\partial}{\partial r}  
  \left( r D \frac{\partial B}{\partial r} \right) , 
\label{eq-diffusion}
\end{equation}
where $t$ is time, and $r$ is the distance from the $z$-axis. 

The turbulent diffusivity $D$ is suppressed in a magnetic field 
exceeding an energy equipartition value $B_e$ 
\citep{1994AN....315..157K}, where $B_e \simeq 400$ G for typical 
parameters of the solar photosphere \citep{1997ApJ...485..398P}. 
It follows that $D$ is a decreasing function of $B$, 
although the exact functional dependence remains uncertain. 
Below we investigate several choices for $D=D(B)$, which yield 
analytical solutions. 

As previously \citep{2015ApJ...800..130L}, 
we choose dimensionless units so that $B_e = r_0 = D_0 = 1$, 
where $D_0 = D(B_e)$ and $r_0$ is the initial radius of the fluxtube. 
Consequently the time is measured in units of $r_0^2 / D_0$. 
The initial value problem is specified by the dimensionless magnetic field 
profile $B(r, 0)$ at $t=0$, where 
\begin{equation}
  B(0, 0) = B_0 
\label{eq-B-0-0}
\end{equation}
is the maximum field within the spot. 
Another important parameter is the total magnetic flux of the spot 
\begin{equation}
  \Phi_0 = 2 \pi \int_0^{\infty} B r dr . 
\label{eq-Phi0-def}
\end{equation}

An initially present sunspot corresponds to $B_0 > 1$ 
(or $B_0 > B_e$ in dimensional units). In the following we define 
the dimensionless fluxtube radius $r_e(t)$ by the condition that 
\begin{equation}
  B(r_e, t) = 1 
\label{eq-edge}
\end{equation}
at its edge. 
Our choice of the length scale $r_0$ to be the initial radius of the spot 
means that 
\begin{equation}
  r_e (0) = 1 , 
\label{eq-r=1}
\end{equation}
and it follows that 
\begin{equation}
  B(1, 0) = 1 . 
\label{eq-B-1-0}
\end{equation}
Finally, we define the spot decay time $T$ by the condition 
\begin{equation}
  r_e(T) = 0 , 
\label{eq-T-def}
\end{equation}
which is equivalent to 
\begin{equation}
  B(0,T) = 1 . 
\label{eq-B-0-T}
\end{equation}

\section{Exact self-similar solutions for nonlinear diffusion of a magnetic flux tube}

Similarity solutions to partial differential equations 
have a large number of applications \citep[e.g.][]{barenblatt-1996}. 
In particular self-similar solutions to nonlinear diffusion equations 
have been considered for several forms of the diffusion coefficient 
\citep[e.g.][]{1990JPhA...23.3681K}. 
These similarity reductions not only lead to exact solutions of specific 
initial-value problems but also serve as intermediate asymptotics that 
approximate solutions of a much larger class of problems. 
Here we consider a similarity reduction of the nonlinear 
two-dimensional diffusion equation (\ref{eq-diffusion}), 
assuming that the dimensionless diffusion coefficient has a power-law form: 
\begin{equation}
  D(B) = B^{-\nu} , 
\label{eq-plaw-dcoeff}
\end{equation}
where we take $\nu > 0$ in order to model the supression 
of magnetic diffusivity in the strong magnetic field of a sunspot.

The self-similar solution to equation (\ref{eq-diffusion}), 
which satisfies the flux conservation condition 
$\Phi_0 = \mbox{const}$, is known to take the form 
\begin{equation}
  B(r, t) = (t_0+t)^{-1/(1-\nu)} \phi(\xi) , \quad 
  \xi = r (t_0+t)^{-1/2(1-\nu)}
%\label{eq-}
\end{equation}
with $t_0 = \mbox{const}$ \citep{barenblatt-1952,pattle-1959}. 
On substituting this form into equation (\ref{eq-diffusion}), 
solving for $\phi(\xi)$, and using equation (\ref{eq-Phi0-def}) 
to specify an integration constant, we get the following 
expression for an evolving field profile: 
\begin{equation}
  B(r, t) = (t_0+t)^{-1/(1-\nu)} \left[ 
  \left( \frac{4 \pi}{\Phi_0} \right)^{\nu/(1-\nu)} + 
  \frac{\nu}{4 (1-\nu)} (t_0+t)^{-1/(1-\nu)} r^2 
  \right]^{-1/\nu} . 
\label{eq-essim-solution}
\end{equation}
It follows that the maximum field (at $r=0$) is given by 
\begin{equation}
  B(0, t) = 
  \left( \frac{\Phi_0}{4 \pi (t_0+t)} \right)^{1/(1-\nu)} , 
\label{eq-B-0-t-selfsim}
\end{equation}
and so 
\begin{equation}
  B_0 = 
  \left( \frac{\Phi_0}{4 \pi t_0} \right)^{1/(1-\nu)} . 
%\label{eq-}
\end{equation}
Now equation (\ref{eq-B-1-0}) yields the magnetic flux 
\begin{equation}
  \Phi_0 = 
  \frac{\pi \nu}{(1-\nu)} 
  \frac{B_0}{B_0^{\nu}-1} . 
\label{eq-Phi0-selfsim}
\end{equation}

To find the decay time $T$ for a given initial magnetic field $B_0$, 
we express the parameter $t_0$ in terms of $B_0$ and substitute 
the resulting expression into equation (\ref{eq-B-0-t-selfsim}). 
On solving equation (\ref{eq-B-0-T}) for $T$ 
and using equation (\ref{eq-Phi0-selfsim}) to eliminate $\Phi_0$, 
we obtain the sunspot decay time in terms of the parameters $B_0$ and $\nu$: 
\begin{equation}
  T = \frac{\nu}{4 (1-\nu)} 
  \frac{B_0 - B_0^{\nu}}{B_0^{\nu}-1} . 
\label{eq-T-selfsim}
\end{equation}
In the limit $\nu \to 0$, equation (\ref{eq-T-selfsim}) reduces
to the expression for the case of linear diffusion: 
\begin{equation}
  \left. T \right|_{\nu=0} = 
  \frac{B_0-1}{4\ln B_0} .  
%\label{}
\end{equation}

Note that the self-similar solution has a curious feature: 
equation (\ref{eq-T-selfsim}) predicts that $T \to 1/4$ as $B_0 \to 1$ 
(for any value of $\nu$), whereas it is physically obvious that $T=0$ 
when $B_0 = 1$. This singular limit behavior is related to the fact that 
the magnetic flux $\Phi_0 \to \infty$ as $B_0 \to 1$. 
In practice this does not cause any problems since we always 
assume the initial field $B_0 > 1$ in order to model a sunspot. 

The self-similar solution above is applicable only for $\nu < 1$ 
since the flux integral in equation (\ref{eq-Phi0-def}) diverges 
for larger values of $\nu$, which physically corresponds to 
an instanteneous flux transfer to infinity \citep{landau-lifshitz-1987}. 
More generally, $\nu < 2/N$ is required to avoid the divergence 
of the flux integral for diffusion in $N$ dimensions. 
Mathematical issues of existence and uniqueness of solutions 
were analyzed by \citet{brezis-friedman-1983}. 

While solutions for $\nu \ge 1$ formally violate the total flux conservation, 
they are mathematically correct and may provide a useful local 
description of nonlinear diffusion. 
As an illustration, consider the case $\nu = 1$. It is straightforward 
to derive a separable solution to equation (\ref{eq-diffusion}). 
By assuming $B(r, t) = (T-t) F(r)$, we reduce the problem to 
a second-order ordinary differential equation for the spatial part $F(r)$. 
The solution is as follows: 
\begin{equation}
  F(r) = 
  \frac{8 c_0}{(c_0 + r^2)^2} , 
\label{eq-Fnu1}
\end{equation}
where $c_0$ is an integration constant, and the other integration constant 
is determined by the requirement that $F$ be finite at $r=0$. 
In a different context, equation (\ref{eq-Fnu1}) in a particular case $c_0 = 1$ 
was given by \citet{1995PhRvL..74.1056R}. 

On using equations (\ref{eq-B-0-0}) and (\ref{eq-Phi0-def}) 
to express $T$ and $c_0$ in terms of $B_0$ and $\Phi_0$, we obtain 
\begin{equation}
  \frac{B(r, t)}{B_0} = 
  \left( 1 + \frac{\pi B_0 r^2}{\Phi_0} \right)^{-2} 
  \left( 1 - \frac{8 \pi t}{\Phi_0} \right) . 
%\label{eq-}
\end{equation}
Here $\Phi_0 = \Phi(0)$ is the initial magnetic flux of the spot 
(at $t=0$), which decreases with time according to 
\begin{equation}
  \Phi(t) = \Phi_0 - 8 \pi t . 
%\label{eq-}
\end{equation}
The localized magnetic field profile is seen to shrink with time 
until the spot vanishes at 
\begin{equation}
  T = \frac{\Phi_0}{8 \pi} . 
\label{eq-T-sfast}
\end{equation}
The termination of the process in a finite time was referred to by
\citet{1995PhRvL..74.1056R} as ``superfast'' diffusion.
This unusual feature of the solution is related to the singular behavior 
of $D(B)=1/B$ as $B \to 0$: in sharp contrast to the solution of a linear problem, 
the continuity flux 
\begin{equation}
\begin{split}
{\cal F}&=-2 \pi r D \frac{\partial B}{\partial r} \\
&=\frac{8\pi^2 r^2}{\Phi_0+\pi B_0 r^2}
\end{split}
\end{equation} 
is independent of time and approaches a constant value, 
${\cal F}\rightarrow 8\pi/B_0$,
as $r \to \infty$, which results in ``flux suction at infinity'' 
\citep{1995PhRvL..74.1056R}. Physically, because $B \to 0$ as $t \to T$, 
the diffusivity $D \to \infty$. Consequently the diffusion time scale 
$1/D \to 0$, and so a diffusive description breaks down as $t \to T$. 

\section{Solutions for the case of a moderately suppressed magnetic diffusivity within a spot}

We now consider a different model for the diffusivity suppression (quenching) 
within a spot, which complements the analysis by \citet{2015ApJ...800..130L}. 
We assume that the evolution of the magnetic field is described by the nonlinear 
two-dimensional diffusion equation (\ref{eq-diffusion}) 
with a step dependence in the dimensionless diffusion coefficient: 
\begin{equation}
  D = 1, \quad B < 1 
\label{eq-dc-step1}
\end{equation}
and 
\begin{equation}
  D = 1-\epsilon, \quad B > 1 . 
\label{eq-dc-step2}
\end{equation}
\citet{2015ApJ...800..130L} considered the case $\epsilon=1$, 
which corresponds to a very strong suppression of turbulent diffusivity 
within the spot. A more realistic model would correspond to a less severe 
turbulent suppression of the diffusivity within the spot. 
Therefore we generalize our nonlinear diffusion model to incorporate the effect 
of a non-vanishing diffusivity within a sunspot on the rate of its disintegration. 

To obtain an analytical solution for the case of 
a moderate diffusivity suppression within the spot, 
we use the heat-balance integral method \citep{goodman-1958}, 
which proved to yield accurate approximations in problems 
of nonlinear diffusion \citep{hill-dewynne-1987,barenblatt-1996}. 
The basic idea is to require that an approximate solution satisfy 
an integral of a nonlinear equation rather than the equation itself. 
We apply the method to describe the turbulent erosion of 
a sunspot, modeled as a cylindrically symmetric fluxtube. 

We assume that $0<\epsilon \ll 1$ and seek an approximate solution of the form 
\begin{equation}
  B(r, t) = f(t) \exp [ -g(t) r^2 ] . 
\label{eq-B-form}
\end{equation}
The fluxtube radius $r_e(t)$ is defined by equation (\ref{eq-edge}). 
Consequently we have 
\begin{equation}
  f = \exp ( g r_e^2 ) . 
\label{eq-fg-rel}
\end{equation}
Integration of equation (\ref{eq-diffusion}) 
over $r$ from $0$ to $\infty$ and 
substitution of the self-similar form (\ref{eq-B-form}) 
into the resulting equation yields 
\begin{equation}
  \frac{d}{dt} \left[ \frac{\exp ( g r_e^2 )}{g} \right] = 
  4 \epsilon g r_e^2 , 
\label{eq-re-ode}
\end{equation}
where equation (\ref{eq-edge}) was used to simplify the right-hand side, 
and equation (\ref{eq-fg-rel}) was used to eliminate $f(t)$. 

In the linear case $\epsilon=0$, the solution of the initial 
value problem with a Gaussian profile at $t=0$ is given by 
\begin{equation}
  B(r, t) = 
  \frac{B_0}{1 + 4 t \ln B_0} 
  \exp \left( - \frac{r^2 \ln B_0}{1 + 4 t \ln B_0} \right) . 
%\label{eq-}
\end{equation}
Here the parameters of an evolving Gaussian profile 
are chosen to satify the initial conditions, 
given by equations (\ref{eq-B-0-0}) and (\ref{eq-B-1-0}). 

The decay time, defined by equations (\ref{eq-T-def}) and (\ref{eq-B-0-T}), 
is easily shown to be 
\begin{equation}
  \left. T \right|_{\epsilon=0} = 
  \frac{B_0-1}{4\ln B_0} , 
\label{eq-T-lin}
\end{equation}
where, as previously, we assume $B_0 > 1$ in order to exclude solutions 
with an infinite magnetic flux. Note for clarity that, if $B_0>e$, 
we have $dr_e/dt>0$ at $t=0$ in the linear solution, and so 
diffusion causes the fluxtube radius to increase until it reaches 
a maximum at $t=(B_0/e-1)/4\ln B_0$ and then to decrease. 

Motivated by the form of the linear solution, we substitute 
\begin{equation}
  g(t) = 
  \frac{\ln B_0}{1 + 4 t \ln B_0} 
\label{eq-g-fun}
\end{equation}
into the approximate equation (\ref{eq-B-form}) describing 
a weakly nonlinear case $0 < \epsilon \ll 1$. Thus 
equation (\ref{eq-re-ode}) becomes an ordinary differential 
equation for the fluxtube radius $r_e(t)$, 
which should be solved subject to the initial condition given 
by equation (\ref{eq-r=1}). Equation (\ref{eq-T-def}) then 
yields the spot decay time $T$. 

Keeping in mind that $\epsilon$ is assumed to be small, 
we solve equation (\ref{eq-re-ode}) by iteration. 
On substituting a simple linear function 
\begin{equation}
  r_e^2 \approx 1 - \frac{t}{T} 
\label{eq-re2-lin}
\end{equation}
into the right-hand side of equation (\ref{eq-re-ode}) and integrating 
from $0$ to $t$, we get an approximate analytical expression for $r_e^2$: 
\begin{equation}
  r_e^2 \approx 
  \frac{1 + 4 t \ln B_0}{\ln B_0} 
  \ln \left\{ 
  \frac{B_0}{1 + 4 t \ln B_0} + 
  \epsilon \frac{\ln B_0}{1 + 4 t \ln B_0} \left[ 
  \left( 1 + \frac{1}{4 T \ln B_0} \right) 
  \ln \left( 1 + 4 t \ln B_0 \right) - 
  \frac{t}{T}
  \right] 
  \right\} . 
\label{eq-re-approx}
\end{equation}
On setting $r_e^2(T)=0$, we obtain an algebraic equation 
for the decay time $T$: 
 \begin{equation}
   \frac{B_0}{1 + 4 T \ln B_0} + 
  \epsilon \frac{\ln B_0}{1 + 4 T \ln B_0} \left[ 
  \left( 1 + \frac{1}{4 T \ln B_0} \right) 
  \ln \left( 1 + 4 T \ln B_0 \right) - 1 
  \right] 
  = 1 . 
\label{eq-T-algebr}
\end{equation}

Alternatively, integration of equation (\ref{eq-re-ode}) over $t$ 
from $0$ to $T$ yields an expression for $T$, which makes clear that 
$\epsilon > 0$ leads to a slower decay: 
\begin{equation}
  T = \frac{B_0-1}{4\ln B_0} + 
  \epsilon \int_0^T g r_e^2 dt . 
\label{eq-T-integral}
\end{equation}
Now substitution of equations (\ref{eq-g-fun}) and (\ref{eq-re2-lin}) 
into the right-hand side of equation (\ref{eq-T-integral}) 
yields equation (\ref{eq-T-algebr}). 

To solve equation (\ref{eq-T-algebr}) in the case of a small $\epsilon$, 
we replace $T$ by $T|_{\epsilon=0}$ from equation (\ref{eq-T-lin}) 
in all terms containing $\epsilon$. The result is as follows: 
\begin{equation}
  T \approx \frac{B_0-1}{4 \ln B_0} + 
  \frac{\epsilon}{4} \left[ 
  \left( 1 + \frac{1}{B_0-1} \right) \ln B_0 - 1 
  \right] . 
\label{eq-T-approx}
\end{equation}
For a fixed $\epsilon \ge 0$, 
$T=T(B_0)$ is an increasing function of $B_0$. 

%{\bf 
We note that the assumed form for the magnetic diffusivity $D$, given by 
equations~(\ref{eq-dc-step1}) and~(\ref{eq-dc-step2}), is finite 
in the limiting case $B \to \infty$, whilst the real magnetic diffusivity is
expected to vanish in this limit. However, 
since $B(r.t)$ remains finite in our solution for any $t>0$, 
the behavior of $D$ as $B \to \infty$ is irrelevant. Our assumption of a 
suppressed but nonzero diffusivity inside the spot is physically meaningful 
as long as the magnetic field $B$ remains finite, as it does in our solution. 
We have previously considered a solution for $B(r,t)$ valid when 
the diffusivity vanishes within the spot \citep{2015ApJ...800..130L}, 
which is accurate for a very strong magnetic field $B_0$ in the spot. 
The new solution presented here quantifies the 
effects of a nonvanishing $D$. 
%}

\section{Numerical solution}

To quantify the accuracy of the analytical results, 
Equation~(\ref{eq-diffusion}) is solved numerically. 
We use an explicit scheme which maps the region $[0,\infty]$ in radius $r$ 
to the region $[0,1]$ in a transformed independent variable $x$, 
as explained in the Appendix. The approach allows an exact boundary condition 
to be imposed at $r=\infty$, which provides more accurate solutions than
an approximate boundary condition at a finite radius. We present solutions with a grid 
spacing $h=0.01$ in $x$ and with a time step one quarter of the stability limit identified 
in the Appendix. 
%The numerical solutions presented here conserve flux to within 
%$0.11\%$  (in the worst case), except where noted.

\subsection{Solutions for the exact self-similar case}

To test the numerical method, and to illustrate the properties of the analytic solutions
presented in Section~3, we consider solution with the power-law form for the diffusion
coefficient, equation~(\ref{eq-plaw-dcoeff}), which admits self-similar solutions.

First we consider a flux-conserving case ($\nu=0.5$), with $B_0=5$. The solid curves in 
the upper panel of Figure~1 show the analytic magnetic field profile 
$B(r,t)$ given by equation~(\ref{eq-essim-solution}) at the three times 
during the spot evolution $t=0$, $t=\frac{1}{4}T$, and $t=T$, where $T$ is 
the decay time defined by equation (\ref{eq-T-selfsim}). The numerical
solutions are also shown in this panel by dashed curves, but they coincide with the 
solid curves and are not visible. The lower panel shows the absolute error in the 
numerical solution at the times $t=\frac{1}{4}T$ (circles) and $t=T$ (plus signs). The
maximum error is $\approx 2\times 10^{-4}$. The numerical solution conserves flux 
throughout the time evolution ($3.2\times 10^4$ time steps) to within $0.11\%$.

\begin{figure}
\figurenum{1}
\centering
% f1: bprofiles_errors_essim_B0eq5.pdf
\plotone{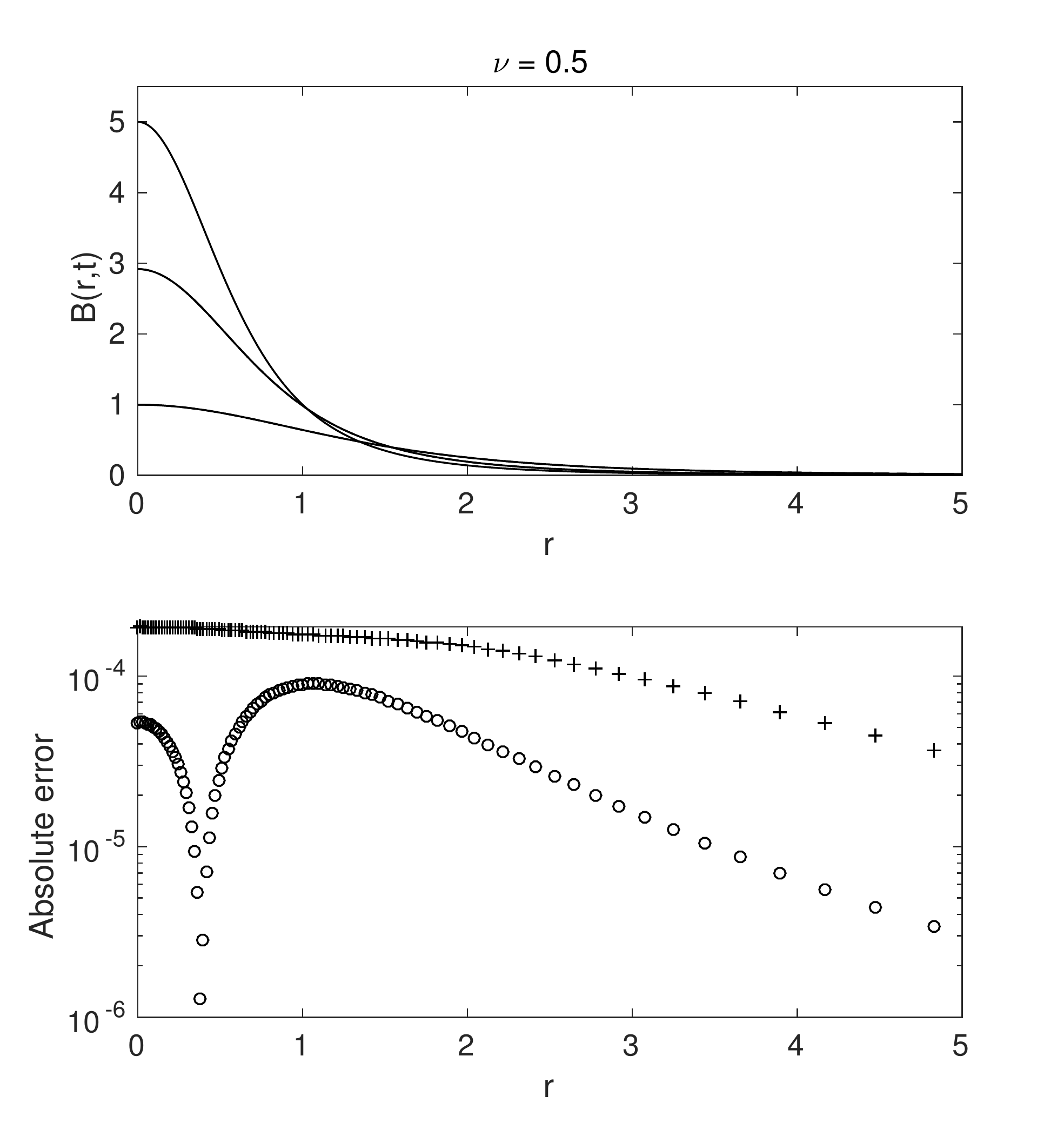}
\caption{Solution to Equation~(\ref{eq-diffusion}) for the case 
$D(B)=B^{-\nu}$ with $\nu=0.5$,
and $B_0=5$. The upper panel shows the analytic (solid) and numerical
(dashed) solutions at $t=0$, $t=\frac{1}{4}T$, and $t=T$. The numerical 
solutions coincide with the analytic solutions and are not visible. The lower panel plots 
the absolute error in the numerical solution at $t=\frac{1}{4}T$ (circles), 
and $t=T$ (plus signs).\label{fig:f1}}
\end{figure}

Second we consider the non-flux conserving case $\nu=1$, 
which exhibits superfast diffusion, i.e.\ vanishing of the spot 
at the time $T$ defined by equation~(\ref{eq-T-sfast}). 
The upper panel of Figure~2 shows 
the analytic and numerical solutions at the three times $t=0$, $t=\frac{1}{4}T$, and 
$t=T$, for the case $B_0=\Phi_0=1$, with a log-linear display used to illustrate 
the behaviour for large $r$. The analytic solutions are the solid curves, and the 
numerical solutions at times $t=\frac{1}{4}T$ and $t=T$ are shown by the 
circles and by the plus signs, respectively. The analytic solution at time $t=T$ is 
identically zero.  The numerical solution is accurate initially,
but becomes inaccurate as $t\rightarrow T$. The reason for the error is that 
the boundary condition at infinity, equation~(\ref{eq-bc2}), does not 
reproduce the behaviour of the continuity flux at large $r$, as shown in
the lower panel. In particular the numerical solution does not maintain a
large positive value of the continuity flux as $r\rightarrow\infty$, which
produces the ``flux suction at infinity''. 
This example demonstrates the need to accurately
represent the boundary conditions at large $r$ in the solutions.

\begin{figure}
\figurenum{2}
\centering
% f2: bprofiles_cflux_sfast.pdf
\plotone{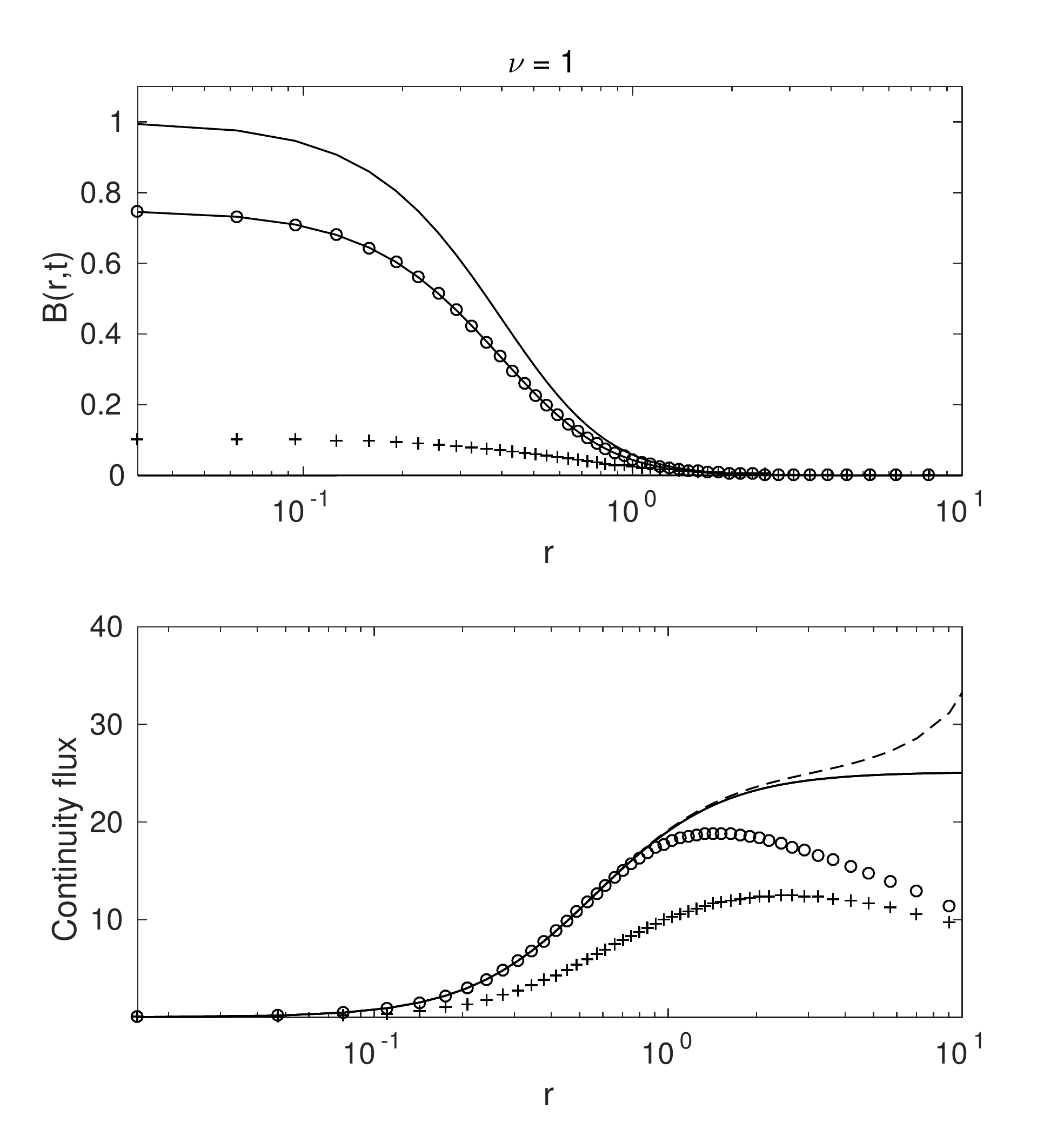}
\caption{Solution to Equation~(\ref{eq-diffusion}) for the case 
$D(B)=B^{-\nu}$ with $\nu=1$, an example of superfast diffusion. 
The 
upper panel shows the field profile $B(r,t)$ at times $t=0$, $t=\frac{1}{4}T$, and $t=T$,
for $B_0=\Phi_0=1$. The analytic solutions are solid curves, and
the numerical 
solutions are shown by circles and plus signs at times $t=\frac{1}{4}T$, and $t=T$.
At time $t=T$ the analytic solution is zero. 
The lower panel plots the continuity flux, which is independent of time in the analytic 
solution (solid curve). The continuity flux in the numerical solution is shown by
a dashed curve ($t=0$), circles ($t=\frac{1}{4}T$), and plus signs
($t=T$). A log-linear display is used to show the behaviour for 
large $r$. \label{fig:f2}}
\end{figure}

\subsection{Solutions for the case of a moderate suppression of $D(B)$ within a spot}

The numerical solution for the case with a diffusion coefficient defined by
equations~(\ref{eq-dc-step1}) and~(\ref{eq-dc-step2}) provides a test of the 
heat-balance integral method results presented in Section~4, and in particular of 
the expression~(\ref{eq-T-approx}) for the lifetime. In this case we do not have an 
exact solution to ensure accuracy, but the numerical solutions 
presented here conserve flux during the time evolution to within $4\times 10^{-4}\%$,
in the worst case.

First we consider the field profile $B(r,t)$ for the heat-balance solution, 
specified by equations~(\ref{eq-B-form}), (\ref{eq-fg-rel}), (\ref{eq-g-fun}), and 
(\ref{eq-re-approx}). Figure~3 compares the heat-balance method profiles
(solid curves) with the numerical solutions (dashed curves) at the three times $t=0$, 
$t=\frac{1}{4}T$, and $t=T$. The solutions assume an 
initial magnetic field strength $B_0=5$. The left panel shows the case $\epsilon=0.25$, 
and the right panel shows the case $\epsilon=0.5$. For the smaller value of $\epsilon$ the 
heat-balance integral method provides a good approximation to the field profile 
throughout the evolution. The method is somewhat less accurate for the larger value 
of $\epsilon$.

\begin{figure}
\figurenum{3}
% f3a: bprofiles_hbal_B0eq5_EPSeq025.pdf, f3b: bprofiles_hbal_B0eq5_EPSeq05.pdf
\plottwo{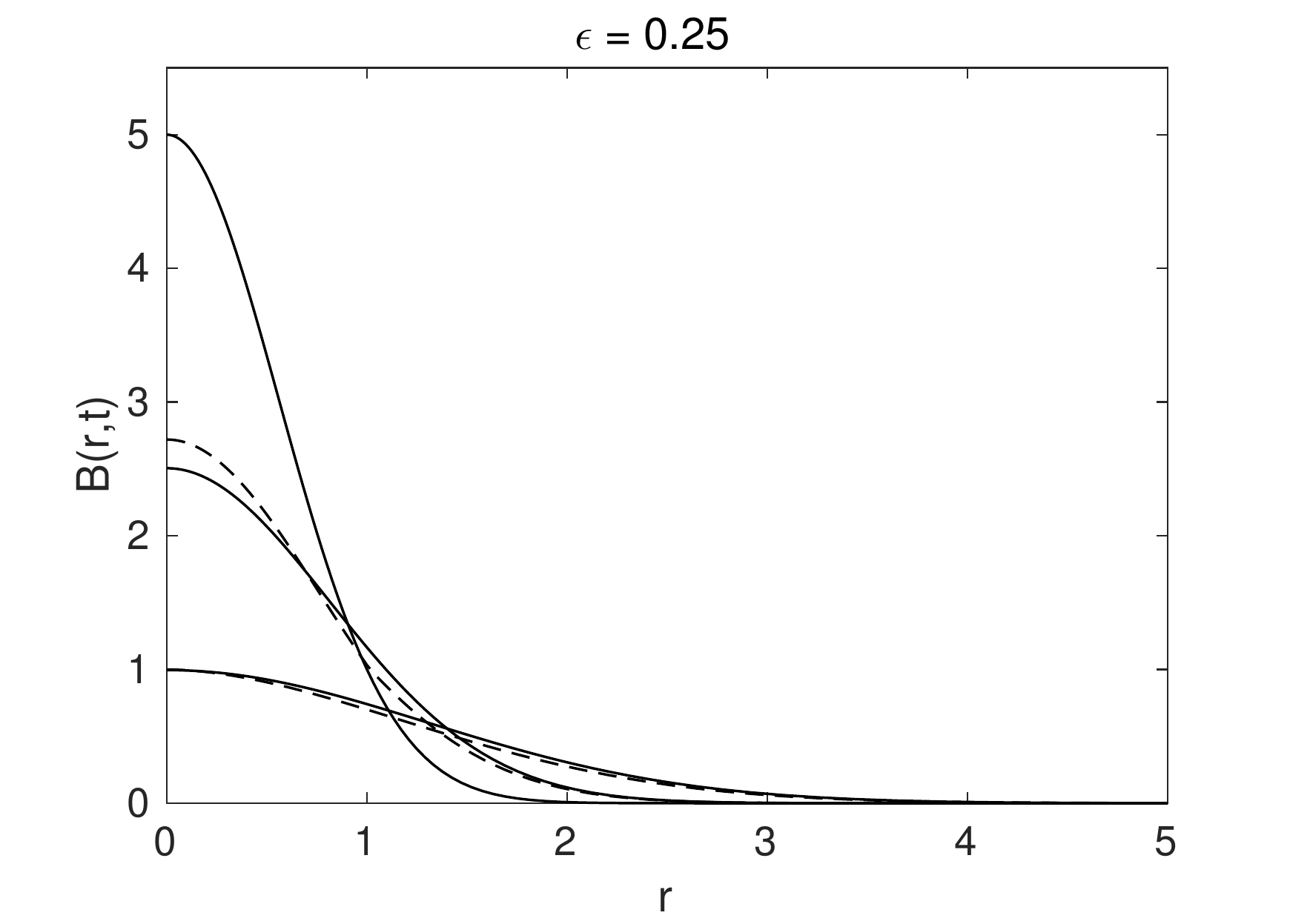}{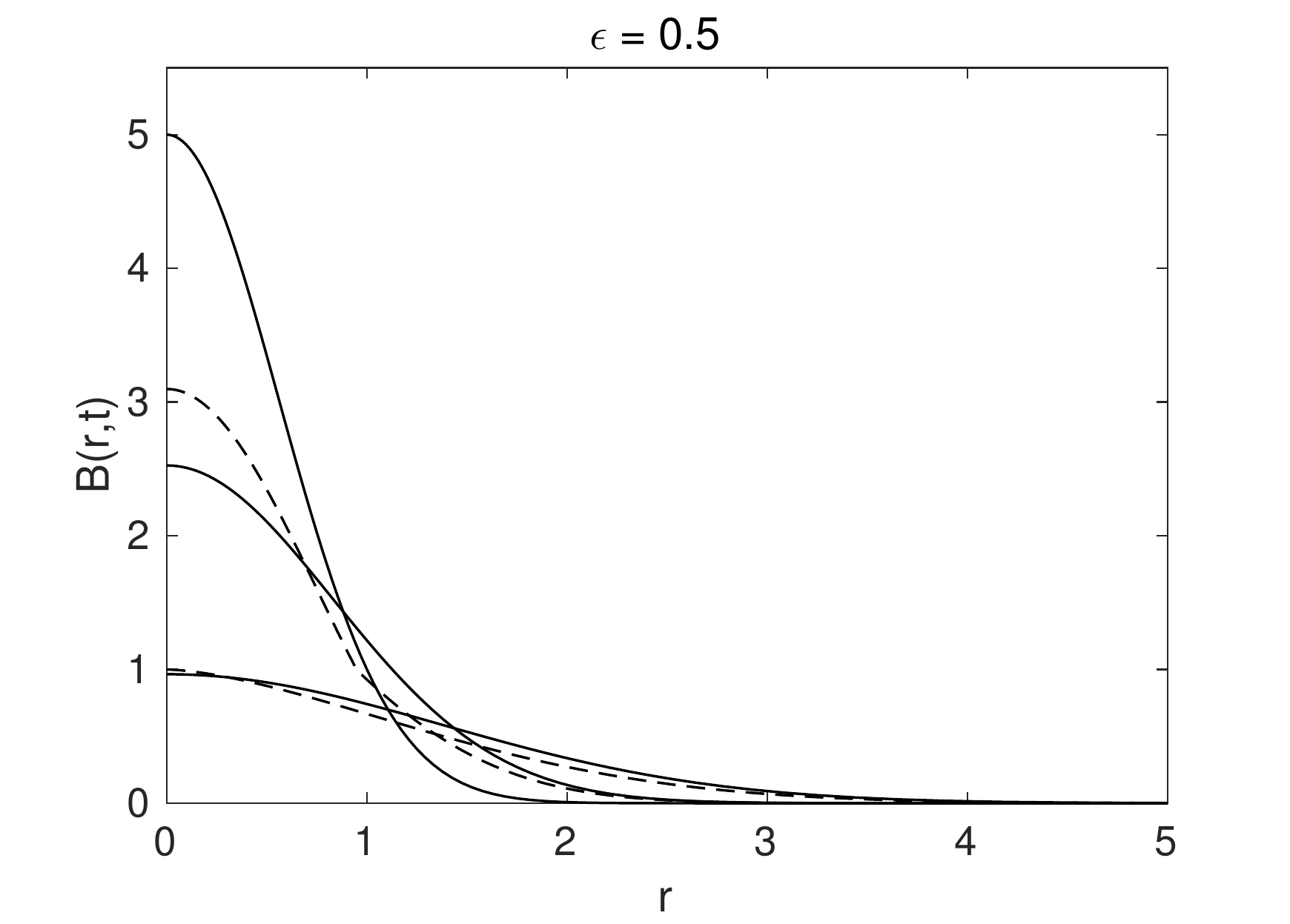}
\caption{Comparison of the heat-balance integral solution (solid) and the numerical
solution (dashed) at $t=0$, $t=\frac{1}{4}T$, and $t=T$, for a spot with $B_0=5$.
The left panel is the case $\epsilon=0.25$, and the right panel is the case 
$\epsilon=0.5$.\label{fig:f3}}
\end{figure}

Figure~4 repeats the display in Figure~1, but shows the case $B_0=10$. The 
accuracy of the field profile obtained with the heat-balance integral method is not 
strongly dependent on the choice of $B_0$.

\begin{figure}
\figurenum{4}
% f4a: bprofiles_hbal_B0eq10_EPSeq025.pdf, f4b: bprofiles_hbal_B0eq10_EPSeq05.pdf
\plottwo{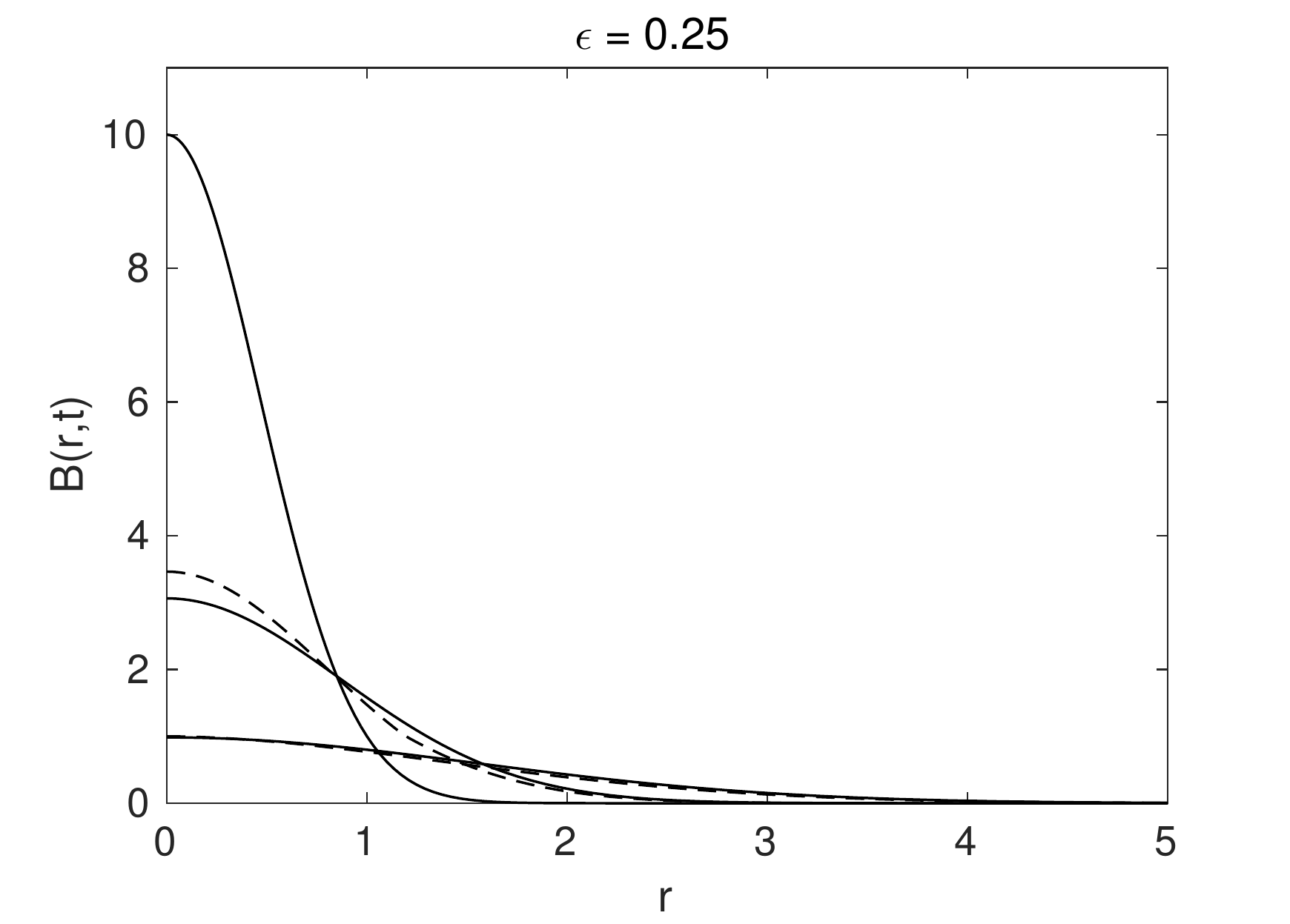}{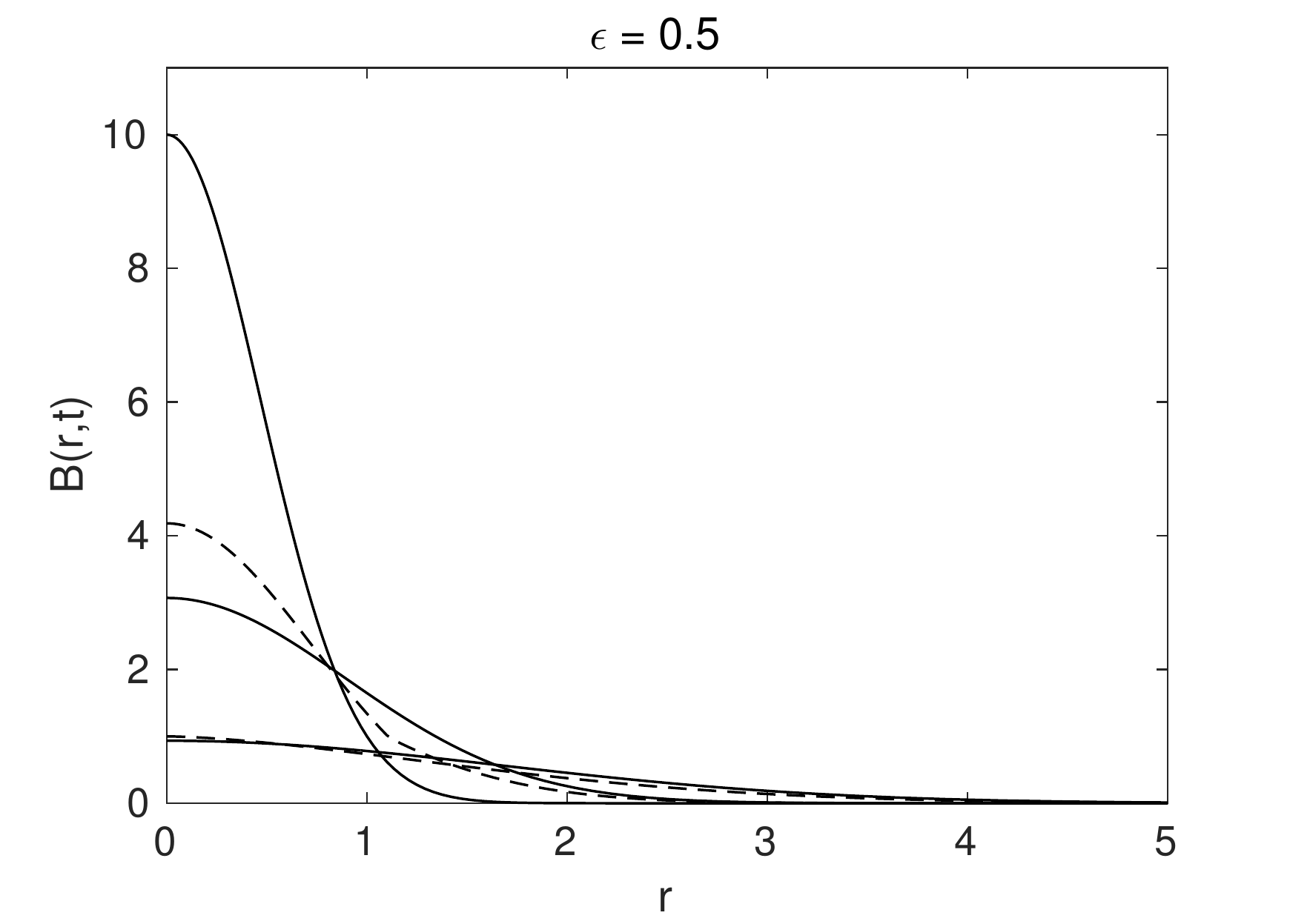}
\caption{Comparison of the heat-balance integral solution (solid) and the numerical
solution (dashed) at $t=0$, $t=\frac{1}{4}T$, and $t=T$, for a spot with $B_0=10$.
The left panel is the case $\epsilon=0.25$, and the right panel is the case 
$\epsilon=0.5$.\label{fig:f4}}
\end{figure}

Second, we consider the accuracy of the heat-balance integral estimate 
for the spot lifetime, equation~(\ref{eq-T-approx}). The estimate
has the form
\begin{equation}
  T \approx T_0 + \epsilon T_1 ,
\label{eq-T-approx-taylor}
\end{equation}
where $T_0$ is the lifetime for the linear case ($\epsilon = 0$), given by 
equation~(\ref{eq-T-lin}). Figure~5 compares 
equation~(\ref{eq-T-approx-taylor}) (solid curves) with the numerical solution
(circles) as a function of $B_0$, for the range $2\leq B_0\leq 10$. 
The upper panel shows $T$ for the three cases $\epsilon=0$, 
$\epsilon =0.25$, and $\epsilon=0.5$ (bottom to top) and the lower 
panel shows $T-T_0$ for the nonlinear cases $\epsilon=0.25$
and $\epsilon=0.5$. Figure~5 demonstrates that the heat-balance method 
provides a good approximation to the lifetime of the spot for the choice
$\epsilon=0.25$, over the range of initial field strengths considered. 
The approximation is worse for the larger value of $\epsilon$, as 
expected.
It is interesting to consider replacing equation~(\ref{eq-T-approx-taylor})
by the $(0,1)$ Pad\'{e} approximant:
\begin{equation}
  T \approx \frac{T_0}{1-\epsilon T_1/T_0}.
\label{eq-T-approx-pade}
\end{equation}
The dotted curves in Figure~5 show the lifetimes given by
equation~(\ref{eq-T-approx-pade}), and the results show that the 
Pad\'{e} expression provides a
better approximation for the lifetime. This might be expected 
based on the final steps in the
derivation of equation~(\ref{eq-T-approx}).

\begin{figure}
\figurenum{5}
\centering
% f5: lifetimes_hbal.pdf
\plotone{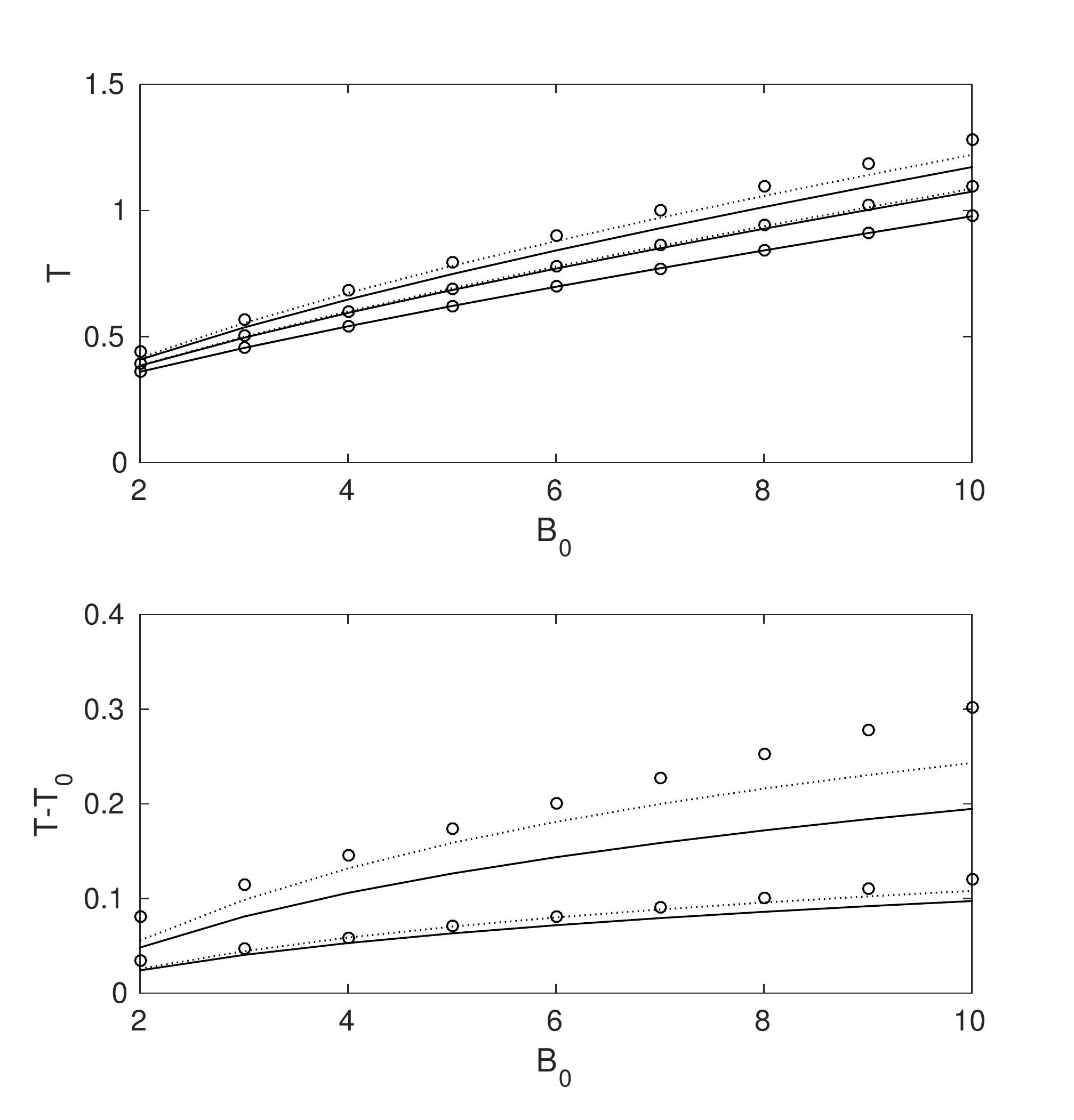}
\caption{The heat-balance integral estimates (solid) for the spot lifetime
and the numerically determined lifetime (circles), plotted as functions of 
$B_0$. The upper panel shows the lifetime $T$ and the lower panel shows
$T-T_0$, where $T_0$ is the linear result ($\epsilon = 0$).
In the upper panel results are shown for $\epsilon=0$, $\epsilon=0.25$,
and $\epsilon=0.5$, and the lower panel shows $\epsilon=0.25$ and 
$\epsilon=0.5$. The dotted curves show the results for the heat-balance integral
method using the Pad\'{e} approximant form for the lifetime.\label{fig:f5}}
\end{figure}

\section{Discussion}

The physical mechanism of sunspot erosion was proposed by \citet{1964ApJ...140.1120S}. 
\citet{1997ApJ...485..398P} introduced a nonlinear diffusion equation 
of a magnetic flux tube as a model of turbulent erosion. 
\citet{1997ApJ...485..398P} identified the maximum magnetic field $B_0$ in a sunspot 
as the key parameter that determines the lifetime $T$ of the spot and derived 
a sunspot decay law due to turbulent erosion. As \citet{2015ApJ...800..130L} 
demonstrated, however, the accuracy of the predictions was limited: 
for instance, \citet{2015ApJ...800..130L} have shown that the sunspot lifetime 
in the model is about a half of that originally predicted. 

We have presented in this paper a further development of the quantitative theory 
of sunspot disintegration by turbulent erosion. 
Our analysis makes it clear that a variety of scalings $T = T(B_0)$ 
are theoretically possible, depending on initial conditions and the dependence 
of the turbulent diffusivity on the magnetic field strength within the spot. 
The results may have implications for the use of the model 
for estimating the magnetic diffusion timescale in starspots  
\citep[e.g.][]{2014ApJ...795...79B,2015ApJ...806..212D,2015A&A...578A.101K}, 
which is an important parameter in stellar dynamo models. 

Experimental studies typically attempt to infer the anomalous, turbulent-driven 
magnetic diffusivity by equating the lifetime of a spot to a theoretical 
diffusion time scale \citep{2014ApJ...795...79B}. 
We expect that the more detailed analytical models we have developed may 
help to improve the accuracy of the procedure. 
More generally, we expect that the new models may allow insight 
into differences and similarities between sunspots and starspots. 
The original formulation of the turbulent erosion model predicted 
a parabolic dependence of the sunspot area on time \citep{1997ApJ...485..398P}. 
Yet the deviations from the parabolic decay are large for any one spot 
\citep{2014SoPh..289.1531G}, motivating the use of continuous linear piecewise 
functions in modeling of the starspot growth and decay \citep{2016ApJ...824..150G}. 
Such an approach is consistent with the more general decay laws predicted 
in our analysis, ultimately controlled by the dependence of 
the magnetic diffusivity on the magnetic field strength within a spot. 
%{\bf 
As pointed out by the referee, however, apparent deviations from the parabolic 
decay law for individual spots can be caused by the difficulty of identifying 
the individual spots within a decaying sunspot group or by the effect of 
a varying external plage field outside the spots 
\citep{1999SoPh..188..315P}.
%}

The turbulent erosion model can be further improved. 
Recall that the solution for $\epsilon=1$ 
in \citet{2015ApJ...800..130L} describes turbulent erosion of a sunspot 
as a moving boundary problem in which the rate of sunspot decay 
is controlled by the inward speed of a current sheet around the spot. 
In other words, the decay rate is determined by the local diffusion 
rate of magnetic field within the sheet, modeled as a tangential 
discontinuity at $r=r_e(t)$. By contrast, the new solution 
given by equations (\ref{eq-re-approx}) and (\ref{eq-T-approx}) 
describes the sunspot decay determined by global magnetic field diffusion, 
and the field discontinuity is ignored in the smooth profile of 
the evolving magnetic field, which should be a reasonable assumption 
if $0 < \epsilon \ll 1$. The diffusive evolution of exact self-similar solutions 
is also a global process. Yet physically the sunspot disintegration rate 
is likely to be influenced by both mechanisms, and so a more accurate solution 
for intermediate values of $\epsilon$ should incorporate 
both local and global diffusion processes by considering 
a more general initial field profile in the spot and more realistic 
diffusivity dependence on the magnetic field strength. 
More detailed models of spot decay should also quantify the effects of 
flux cancellation, caused by photospheric magnetic reconnection 
\citep[e.g.][]{2015JKAS...48..187L}. 
Application of the model to the data on sunspot and starspot decay 
may shed light on the physics of turbulent diffusion 
in magnetized astrophysical plasmas. 

\acknowledgments
An anonymous referee’s comments and suggestions are gratefully acknowledged. 

\appendix

\section{Numerical method}

\citet{2015ApJ...800..130L} numerically solved
Equation~(\ref{eq-diffusion}) using a Crank--Nicolson scheme. The 
method imposed a boundary condition at a finite outer boundary
$r=r_m$ which allowed flux transport across the boundary, using
one-sided spatial derivatives. This approach was an improvement
over the assumption of zero flux at an outer radius used by 
\citet{1997ApJ...485..398P}, but it was still a source of error 
for the long time integrations necessary to determine the lifetimes
for the model spots. Here we use a simpler explicit scheme, but
with a transformation of the infinite $r$-domain to a finite
domain, to allow an exact outer boundary condition.

The change of independent variable 
\begin{equation}
r=\tan \left(\frac{\pi}{2}x\right)
\label{eq-xtor}
\end{equation}
maps $0\leq r\leq\infty$ to $0\leq x\leq 1$. Transforming the 
derivatives in equation~(\ref{eq-diffusion}) gives
\begin{equation}
  \frac{\partial B}{\partial t} = 
  \frac{4}{\pi^2r(1+r^2)} \frac{\partial}{\partial x}  
  \left(\frac{r}{1+r^2} D \frac{\partial B}{\partial x} \right) . 
\label{eq-diff-xr}
\end{equation}

We consider solution of equation~(\ref{eq-diff-xr}) on a uniformly spaced
grid in $x$ defined by
$x_j=(j-1)h$, with $j=1,2,\dots,L$, where $h=1/(L-1)$ is the grid spacing.
Similarly we consider discrete times $t_n=(n-1)\tau$, with $n=1,2,\dots$, where
$\tau$ is a constant time step. 
A suitable forward-time, centred-space (FTCS) discretisation of 
equation~(\ref{eq-diff-xr}) is~\citep{1992nrca.book.....P}:
\begin{equation}
 \frac{B_{j}^{n+1}-B_{j}^n}{\tau}\approx
   \frac{4}{\pi^2r_j(1+r_j^2)h}\left(
\frac{r_{j+\frac{1}{2}}}{1+r_{j+\frac{1}{2}}^2} D_{j+\frac{1}{2}}^{n}
    \frac{B_{j+1}^{n}-B_j^{n}}{h}
 -\frac{r_{j-\frac{1}{2}}}{1+r_{j-\frac{1}{2}}^2} D_{j-\frac{1}{2}}^{n}
    \frac{B_{j}^{n}-B_{j-1}^{n}}{h}
\right) ,
 \label{eq-diff-xr-ftcs}
\end{equation}
where $B_j^n=B(x_j,t_n)$ and $D_j^n=D(B_j^n)$, and where
$r_j=\tan\left(\frac{\pi}{2}x_j\right)$. The diffusion coefficients at
intermediate grid points may be approximated by
\begin{equation}
D_{j\pm \frac{1}{2}}\approx D_{j\pm}=\frac{1}{2}\left(D_j^n+D_{j\pm 1}^n\right).
\label{eq-dcoeff-approx}
\end{equation}
Equations~(\ref{eq-diff-xr-ftcs}) and~(\ref{eq-dcoeff-approx}) give 
the numerical scheme
\begin{equation}
 B_{j}^{n+1} = B_{j}^n
   +\frac{2}{\pi^2r_j(1+r_j^2)}\frac{\tau}{h^2}\left[
\frac{r_{j+\frac{1}{2}}}{1+r_{j+\frac{1}{2}}^2} D_{j+}^{n}
    \left(B_{j+1}^{n}-B_j^{n}\right)
 -\frac{r_{j-\frac{1}{2}}}{1+r_{j-\frac{1}{2}}^2} D_{j-}^{n}
    \left(B_{j}^{n}-B_{j-1}^{n}\right)\right] .
     \label{eq-numerical-scheme}
\end{equation}

Equation~(\ref{eq-numerical-scheme}) is an explicit prescription for 
time evolution at grid locations $j=2,3,\dots ,L-1$. For $j=1$ ($r=0$) the 
physical boundary condition is
\begin{equation}
\left.\frac{\partial B}{\partial r}\right|_{r=0}=
  \left.\frac{2}{\pi (1+r^2)}\frac{\partial B}{\partial x}\right|_{r=x=0}=0
  \label{eq-bc1-1}
  \end{equation}
  for all times.
 Equation~(\ref{eq-bc1-1}) may be enforced using a one-sided (forward) 
 approximation to the derivative with respect to $x$ at $x=0$:
 \begin{equation}
  \left.\frac{\partial B}{\partial x}\right|_{x=0, t_{n+1}}\approx
 \frac{-3B_1^{n+1}+4B_2^{n+1}-B_3^{n+1}}{2h}
 \end{equation}
 giving the update for the grid point $j=1$:
\begin{equation}
B_1^{n+1}=\frac{1}{3}\left(4B_2^{n+1}-B_3^{n+1}\right)  .
\label{eq-bc1-2}
\end{equation}
This approach requires that the initial conditions satisfy equation~(\ref{eq-bc1-2}), 
i.e.\ $B_1^1=\frac{1}{3}\left(4B_2^1-B_3^1\right)$.

For $j=L$, corresponding to $r=\infty$, we use the update
\begin{equation}
B_L^{n+1}=B_L^{n} ,
\label{eq-bc2}
\end{equation}
which together with the initial condition $B_L^1=0$ ensures that the field 
is zero at infinity.

Under the transformation~(\ref{eq-xtor}) the total magnetic flux, defined
by Equation~(\ref{eq-Phi0-def}), becomes:
\begin{equation}
\Phi_0=\pi^2\int_{0}^1 r(x)\left[1+r(x)^2\right]B[r(x),t)\,dx .
\label{eq-Phi0-def-x1}
\end{equation}
Equation~(\ref{eq-Phi0-def-x1}) may be evaluated in our discrete version of
the problem using the trapezoidal rule~\citep{1992nrca.book.....P}:
\begin{equation}
\Phi^n_0=\pi^2h\sum_{j=1}^{L}w_jr_j\left(1+r_j^2\right)B_j^n ,
\label{eq-Phi0-def-x2}
\end{equation}
where $w_j=1$ for $j=2,\dots ,L-1$, and $w_1=w_L=\frac{1}{2}$. 
Equation~(\ref{eq-Phi0-def-x2}) is used to check that flux is approximately
conserved by the numerical solution.

A simple estimate of the stability condition for the
method may be made as follows. We expect that an FTCS discretisation of 
equation~(\ref{eq-diffusion}) is stable at a given time step subject 
to~\citep{1992nrca.book.....P}:
\begin{equation}
\tau^n\leq \min_j \frac{\left(\Delta r_j\right)^2}{4D_j^n} ,
\label{eq-ftcs-stability1}
\end{equation}
where $\Delta r_j$ is the grid spacing. (Strictly this requires a 
uniform grid in $r$.) From equation~(\ref{eq-xtor}) we have 
$\Delta r_j\approx \frac{\pi}{2}\left(1+r_j^2\right) h$ so 
equation~(\ref{eq-ftcs-stability1}) becomes
\begin{equation}
\tau^n\leq \frac{\pi^2h^2}{16}\min_j \frac{\left(1+ r_j^2\right)^2}{D_j^n} .
\label{eq-ftcs-stability2}
\end{equation}
In the case of the model with a step dependence of $D$
on $B$ (Section~4) we can take
\begin{equation}
\tau\leq \frac{\pi^2h^2}{16}\left(1+ r_1^2\right)^2.
\label{eq-ftcs-stability-hbal}
\end{equation}
Numerical experimentation suggests that equation~(\ref{eq-ftcs-stability-hbal}) 
provides a good estimate for the actual stability constraint.

\end{document}